\documentclass[prl, a4paper, twocolumn,amssymb,superscriptaddress, showpacs]{revtex4}
\usepackage{dsfont, amsmath, verbatim, graphicx, color, ulem}
\usepackage{epsfig}
\usepackage{bbm}
\usepackage[latin1]{inputenc} % Umlaute!

\newcommand{\bra}[1]{\langle #1|}
\newcommand{\ket}[1]{|#1\rangle}
\newcommand{\braket}[2]{\langle #1 | #2 \rangle}
\newcommand{\ketbra}[1]{| #1\rangle \langle #1|}

\newcommand{\be}{\begin{equation}}
\newcommand{\ee}{\end{equation}}
\newcommand{\eea}{\end{eqnarray}}
\newcommand{\bea}{\begin{eqnarray}}

\newcommand{\mean}[1]{\ensuremath{\langle{#1}\rangle}}

\newcommand{\eins}{\openone}
\newcommand{\qed}{\ensuremath{\hfill \Box}}
\newcommand{\WW}{\ensuremath{\mathcal{W}}}

\newcommand{\MM}{\ensuremath{\mathcal{M}}}

\newcommand{\BB}{\ensuremath{\mathcal{B}}}

\newcommand{\VV}{\ensuremath{\mathcal{V}}}
\newcommand{\SSS}{\ensuremath{\mathcal{S}}}
\newcommand{\EE}{\ensuremath{\mathcal{E}}}

\newcommand{\kommentar}[1]{}

\renewcommand{\vr}{\ensuremath{\varrho}}

\newcommand{\forget}[1]{}

\DeclareMathOperator{\tr}{Tr}
\DeclareMathOperator{\real}{Re}
\newcommand{\abs}[1]{|#1|}
\newcommand{\id}{\mathbf{1}}

\begin{document}
\title{Increasing the statistical significance of entanglement 
detection in experiments 
}
\author{Bastian Jungnitsch}
\affiliation{Institut f\"{u}r Quantenoptik und Quanteninformation,
\"{O}sterreichische Akademie der Wissenschaften, Technikerstra{\ss}e
21A, A-6020 Innsbruck, Austria\\}
\author{S\"onke Niekamp}
\affiliation{Institut f\"{u}r Quantenoptik und Quanteninformation,
\"{O}sterreichische Akademie der Wissenschaften, Technikerstra{\ss}e
21A, A-6020 Innsbruck, Austria\\}
\author{Matthias Kleinmann}
\affiliation{Institut f\"{u}r Quantenoptik und Quanteninformation,
\"{O}sterreichische Akademie der Wissenschaften, Technikerstra{\ss}e
21A, A-6020 Innsbruck, Austria\\}
\author{Otfried~G\"{u}hne }
\affiliation{Institut f\"{u}r Quantenoptik und Quanteninformation,
\"{O}sterreichische Akademie der Wissenschaften, Technikerstra{\ss}e
21A, A-6020 Innsbruck, Austria\\} 
\affiliation{Institut f{\"u}r
Theoretische Physik, Universit{\"a}t Innsbruck, Technikerstra{\ss}e
25, A-6020 Innsbruck, Austria\\}
\author{He Lu}
\affiliation{Hefei National Laboratory for Physical Sciences at
Microscale and Department of Modern Physics, University of Science
and Technology of China, Hefei, Anhui 230026, China}
\author{Wei-Bo Gao}
\affiliation{Hefei National Laboratory for Physical Sciences at
Microscale and Department of Modern Physics, University of Science
and Technology of China, Hefei, Anhui 230026, China}
\author{Yu-Ao Chen}
\affiliation{Hefei National Laboratory for Physical Sciences at
Microscale and Department of Modern Physics, University of Science
and Technology of China, Hefei, Anhui 230026, China}
\affiliation{Physikalisches Institut, Ruprecht-Karls-Universit\"{a}t
Heidelberg, Philosophenweg 12, 69120 Heidelberg, Germany}
\author{Zeng-Bing Chen}
\affiliation{Hefei National Laboratory for Physical Sciences at
Microscale and Department of Modern Physics, University of Science
and Technology of China, Hefei, Anhui 230026, China}
\author{Jian-Wei Pan}
\affiliation{Hefei National Laboratory for Physical Sciences at
Microscale and Department of Modern Physics, University of Science
and Technology of China, Hefei, Anhui 230026, China}
\affiliation{Physikalisches Institut, Ruprecht-Karls-Universit\"{a}t
Heidelberg, Philosophenweg 12, 69120 Heidelberg, Germany}

\date{\today}
\begin{abstract}
Entanglement is often verified by a violation of an 
inequality like a Bell inequality or an entanglement witness. Considerable 
effort has been devoted to the optimization of such inequalities in order to 
obtain a high violation.  We demonstrate theoretically and experimentally that 
such an optimization does not necessarily lead to a better
entanglement test, if the statistical error is taken into 
account. Theoretically, we show for different error models 
that reducing the violation of an inequality can improve 
the significance. Experimentally, we observe this phenomenon
in a four-photon experiment, testing the Mermin and Ardehali 
inequality for different levels of noise. Furthermore, we provide 
a way to develop entanglement tests with high statistical significance.
%\emp{, thereby providing a direction to find powerful entanglement tests. 
%We confirm this increase of the statistical significance experimentally}
%in a four-photon experiment, testing the Mermin and Ardehali 
%inequality for different levels of noise. 
\end{abstract}

\maketitle

%%%%%%%%%%%%%%%%%%%%%%%%%%%%%%%%%%%%%%%%%%%%%%%%%%%%%%%%%
{\it Introduction ---}
%%%%%%%%%%%%%%%%%%%%%%%%%%%%%%%%%%%%%%%%%%%%%%%%%%%%%%%%%
Quantum theory is a statistical theory, predicting in general 
only probabilities for experimental results. Consequently, in 
most experiments observing quantum effects, several copies of a 
quantum state are generated and individually measured to 
determine the desired probabilities. As only a finite number 
of states can be generated, this leads to an unavoidable 
statistical error. The particularly low generation rate in 
certain experiments demands a careful statistical treatment;
a fact that is well known from particle physics 
\cite{hepandastro, feldman}.

In quantum information processing, many of today's experiments 
aim at the generation of entanglement, which is considered to 
be a central resource \cite{hororeview,review}. So far, entanglement 
of up to ten qubits has been achieved using trapped ions or photons 
\cite{experiments,luexp}. For 
the experimental verification of entanglement, often inequalities 
for the correlations --- such as Bell inequalities or entanglement 
witnesses --- are used \cite{review}, in which a violation indicates entanglement. The 
maximization of this violation has been investigated in detail, cf.\ 
Refs.~\cite{review,EWoptimization}. In fact, making such inequalities  more 
sensitive is a crucial step in order to allow advanced experiments with more 
particles.

In this paper we demonstrate theoretically and experimentally
that such an optimization does not necessarily lead to a 
better entanglement test, if the statistical nature of quantum 
theory is taken into account. It was already noted \cite{gill} that, when aiming at ruling out local realism, highly entangled 
states do not necessarily deliver a stronger test than weakly 
entangled states, but this does not answer the question which inequality to use 
for a given state and it remains unclear how to apply it to actual error models 
used in experiments. Also, most of the different entanglement detection methods
compared in Ref.~\cite{altepeter} cannot be applied to multiparticle systems.

%In Ref.~\cite{altepeter} different entanglement detection methods for two qubits 
%have been compared, but most of these methods cannot be applied to multiparticle systems. 

Theoretically, we show for different error models that decreasing 
the violation of an inequality can improve the significance. Also, 
we demonstrate this phenomenon in a four-photon experiment, measuring 
the Mermin and the Ardehali inequality. We find that the former inequality leads to 
a higher significance than the latter, despite a lower violation. Finally, 
we discuss the physical origin of this phenomenon and provide methods 
to construct entanglement tests with a high statistical significance. 
%and thus an approach that results
%in more powerful entanglement tests.

%%%%%%%%%%%%%%%%%%%%%%%%%%%%%%%%%%%%%%%%%%%%%%%%%%%%%%%%%%%%%%%%
{\it Statement of the problem ---}
%%%%%%%%%%%%%%%%%%%%%%%%%%%%%%%%%%%%%%%%%%%%%%%%%%%%%%%%%%%%%%%%
A witness $\WW$ is an observable which has a non-negative
expectation value on all separable states (i.e., states which can be written 
as  a mixture of product states, $\vr = \sum_k p_k \ketbra{a_k,b_k}$ with some 
probabilities $p_k$).
Hence, a negative expectation value of a witness signals entanglement. Similarly, 
a Bell inequality $\mean{\BB} \leq C_{\rm lhv}$, where $\BB$ is a sum
of certain correlation terms, holds if the measurement
outcomes can originate from a local hidden variable (LHV) model.
As separable states allow a description by LHV models, a violation
of a Bell inequality implies the presence of entanglement.

In both cases, we define $\VV$ as the violation of the
corresponding inequality. That is, for a witness we have
$\VV(\WW)= - \mean{\WW}$ while for a Bell inequality
$\VV(\BB)=\mean{\BB}-C_{\rm lhv}.$ Then, the significance
of an entanglement test can be defined as
\be
\SSS = \frac{\VV}{\EE}
\label{significancedef}
\ee
where $\EE$ is the statistical error for the experiment.
Clearly, $\EE$ depends on the particular experimental 
implementation and on the error model used. Nevertheless, 
in any experiment $\SSS$ is a well characterized quantity;
its notion is widely used in the literature, when the violation
is expressed in terms of ``standard deviations'', also
in other fields of physics \cite{hepandastro}.

% \emp{The use of this quantity is not only restricted to quantum information. In fact, it extends to other %fields in which the number of detection events is as small as in multipartite quantum information %experiments, such as particle physics and astrophysics \cite{sigHEPandastro}}.

Previously, much effort has been devoted to improving
entanglement tests in order
to achieve a higher violation. For instance, for entanglement
witnesses a mature theory how to optimize witnesses has been developed 
\cite{EWoptimization}. Here, for a given witness $\WW$ one tries to find a
positive operator $P$, %which one can subtract from the witness
such that $\WW'= \WW-P$ is still a witness. In order to have a more 
significant result, however, one can either increase $\VV$ in 
Eq.~(\ref{significancedef}) or decrease $\EE.$ It is a central result of this 
paper that decreasing $\EE$ is often superior.

%%%%%%%%%%%%%%%%%%%%%%%%%%%%%%%%%%%%%%%%%%%%%%%%%%%%%%%%%%
{\it Variance as the error ---}
%%%%%%%%%%%%%%%%%%%%%%%%%%%%%%%%%%%%%%%%%%%%%%%%%%%%%%%%5
Let us first consider a simple model, in which we take
the square root of the variance as the error of a witness,
\be
\EE(\WW) = \Delta(\WW)= \sqrt{\mean{\WW^2}-\mean{\WW}^2}.
\ee
An experimentally relevant model will be discussed below.
This simple model already demonstrates that the standard 
optimization of witnesses is often not 
the appropriate approach to increase the significance:

{\bf Observation.} Let $\vr=\ket{\psi}\bra{\psi}$ be a pure state
detected by the witness $\WW.$ Then, one can always increase the
significance of $\WW$ at the expense of optimality (i.e., by {\it adding}
a positive operator). With this method one can make the significance
arbitrarily large.

Namely, one needs to find a positive observable $P$, 
such that $\ket{\psi}$ is an eigenstate of $\WW'=\WW+P$;
then the error vanishes. Indeed, such a $P$ can be found (see Appendix).

%%%%%%%%%%%%%%%%%%%%%%%%%%%%%%%%%%%%%%%%%%%%%%%%%%%%%%%%%%%%%%%
{\it Multi-photon experiments ---}
%%%%%%%%%%%%%%%%%%%%%%%%%%%%%%%%%%%%%%%%%%%%%%%%%%%%%%%%%
Let us now consider a realistic situation, in which other and more
specific error models are used. As our later implementation
uses multi-photon entanglement, we concentrate on this type of
experiments but our ideas can also be applied to other implementations, 
such as trapped ions. 
%\cite{hartmutpr}.

The basic experimental quantities are the numbers of detection events $n_i$  
of the different detectors $i$. {From} these data, all other quantities such 
as correlations or mean
values of observables are derived.

In the standard error model for photonic experiments 
\cite{luexp,james}, the counts are assumed to be distributed according 
to a Poissonian distribution, whose mean value is given by the 
observed value. That is, for a certain measurement outcome $i$ one sets the
mean value as $\mean{n_i} = n_i$ and the error as
$\EE(n_i) = \sqrt{n_i}$ (being the standard deviation of a Poissonian distribution).
In general, for a function $f = f(n_i)$ 
%that is a function 
of several counts, Gaussian error propagation is applied to obtain the error (see below).

To give an example, consider a two-qubit correlation
\begin{equation}
\MM =
\alpha Z_1  Z_2
+ \beta Z_1 \eins_2
+ \gamma \eins_1 Z_2.
\end{equation}
Here and in the following, $Z_k$ (or $\eins_k$) denotes
the Pauli matrix $\sigma_z$  (or the identity matrix)
acting on the $k$th qubit and tensor product symbols
are omitted.
$\mean{\MM}$ can be determined by measuring in the common eigenbasis
of all three terms in $\MM$, i.e., by projecting onto
 $\ket{00},\ket{01},\ket{10}$
and $\ket{11}$. Repeating this with many copies of the state
will lead to count numbers $n_{kl}$ with $k,l = 0$ or $1$
and to count rates $p_{kl}=n_{kl}/n_{\rm tot},$ where
$n_{\rm tot}= n_{00}+ n_{01} + n_{10}+ n_{11}$ is the
total number of events. The mean value $\mean{\MM}$
can  be written as a linear combination of $p_{kl}$,
namely
$
\mean{\MM} =
\lambda_{00} p_{00} +\lambda_{01} p_{01} +
\lambda_{10} p_{10} +\lambda_{11} p_{11}
$
with
$
\lambda_{00} = \alpha + \beta + \gamma,
\lambda_{01} = -\alpha + \beta - \gamma,
\lambda_{10} = -\alpha -\beta + \gamma,
$
and
$
\lambda_{11} = \alpha - \beta - \gamma
$.
Then, according to Gaussian error propagation, the squared
error is given by \cite{remarkjames}
\be
\EE(\MM)^2 = {
\sum_{k,l}
\Big[
\frac{\partial \mean{\MM}}{\partial{n_{kl}}}
\Big]^2
\EE(n_{kl})^2
}
=
{
\sum_{k,l}
\Big[
\frac{\lambda_{kl}}{n_{\rm tot}}
-
\frac{\mean{\MM}}{n_{\rm tot}}
\Big]^2
n_{kl}.
}
\label{errorformula}
\ee
Let us finally discuss the underlying assumptions of this
error model. The first main assumption is that the $n_{kl}$
are Poisson distributed and their errors are uncorrelated.
This is well motivated by the experimental observations.
Moreover, Gaussian error propagation stems from a Taylor 
expansion of the function $f$. Finally, if one  
interprets  the standard deviation as a confidence interval, 
one tacitly assumes that the distribution is Gaussian, 
as for other distributions the connection is not so direct.
If the number of events for all detectors is sufficiently large (e.g.\ $n_{kl} 
\gtrsim 10$), however, the Poissonian distribution is approximated well by a 
Gaussian distribution.

%%%%%%%%%%%%%%%%%%%%%%%%%%%%%%%%%%%%%%%%%%%%%%%%%%%
{\it Bell inequalities for four particles ---}
%%%%%%%%%%%%%%%%%%%%%%%%%%%%%%%%%%%%%%%%%%%%%%%%%%%5
Let us now discuss the Mermin and Ardehali inequality as
experimentally relevant examples. First, we consider
\begin{align}
\label{eq:Mermin_op}
\BB_{\rm M} = \:&
X_1 X_2 X_3 X_4
-
[X_1 X_2 Y_3 Y_4 +\!\mbox{ perm.} ] + \: Y_1 Y_2 Y_3 Y_4,
\end{align}
where the bracket $[\dots]$ is meant as a sum over all permutations of 
$X_1 X_2 Y_3 Y_4$ that yield distinct operators. For states allowing an LHV 
description, the Mermin inequality $\mean{\BB_{\rm M}} \leq 4$ holds \cite{Mermin}.
We wrote $\BB_{\rm M}$ with the Pauli matrices as observables, since 
they are used later, however, one might replace them by arbitrary dichotomic 
measurements.

\begin{figure}
\includegraphics[width=0.8\columnwidth]{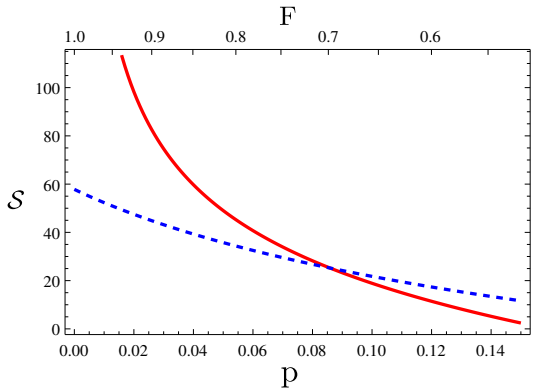}
\caption{\label{fig:bitflipnoise}
Significance $\SSS$  for the Mermin (red, solid) and the 
Ardehali inequality (blue, dashed) for bit-flip noise. 
On the horizontal axes, we show 
%%%
{the bit-flip probability and the
corresponding fidelity with respect to a perfect GHZ state}. We assumed that the experimenter 
prepares 8000 instances of a GHZ state and chooses either
to measure  the eight terms of the Mermin inequality (each 
term with 1000 realizations of the state) or the 16 terms 
of the Ardehali inequality with 500 states per correlation 
term. See text for further details.}
\end{figure}

Second, we consider the Ardehali inequality
$\langle \BB_{\rm A} \rangle \leq  2 \sqrt{2}$
\cite{Ardehali}, where
\begin{align}
\label{eq:Ardehali_op}
\BB_{\rm A}
& =
\left(
X_1 X_2 X_3 A_4 + X_1 X_2 X_3 B_4 \right.
\nonumber
\\
&-[X_1 Y_2 Y_3 A_4 + \mbox{ perm.}] - [X_1 Y_2 Y_3 B_4 + \mbox{ perm.}]
\nonumber
\\
& -[X_1 X_2 Y_3 A_4 + \mbox{ perm.}] + [X_1 X_2 Y_3 B_4 + \mbox{ perm.}]
\nonumber
\\ 
&\left.+ Y_1 Y_2 Y_3 A_4 - Y_1 Y_2 Y_3 B_4
\right)/{\sqrt{2}}.
\end{align}
Here, the sums in square brackets include all distinct permutations
on the first three qubits. We set
$A_4 = \left(X_4 + Y_4\right) / \sqrt{2}$ and
$B_4 = \left(X_4 - Y_4\right) / \sqrt{2}$, 
but again,
%in order to test local realism, 
the observables can remain
arbitrary \cite{remarkidentical}.

The Mermin and Ardehali inequality 
%are designed 
%to 
reveal the non-local correlations of the four-qubit 
GHZ state,
\be
\ket{GHZ_4} = \frac{1}{\sqrt{2}}(\ket{0000}+ \ket{1111}).
\label{ghzdef}
\ee
For this state we have $\mean{\BB_{\rm M}} = \mean{\BB_{\rm A}} = 8.$
As the bound for LHV models for the Ardehali inequality is smaller,
the violation $\VV$ is larger. This may lead to the opinion that 
the Ardehali inequality is ``better'' than the Mermin inequality 
for the state $\ket{GHZ_4}.$

However, this belief is easily shattered, if the significance
$\SSS$ is considered as the relevant figure of merit. This can be seen 
directly 
from 
%the expression of the error in 
Eq.~(\ref{errorformula}).
The GHZ state is an eigenstate for each of the correlation measurements in the 
Mermin inequality (they are so-called stabilizing operators of the GHZ 
state).  Hence, if the Mermin inequality for a perfect GHZ state is measured, 
we have in the last term of Eq.~(\ref{errorformula}) for each case $k,l$ either 
$\lambda_{kl}= \mean{\MM}$ (since the mean value is an eigenvalue) or 
$n_{kl}=0,$ hence 
%the error 
$\EE(\MM)$ vanishes. The Ardehali inequality, 
however, does not contain stabilizer terms and the error remains finite.

For an experimental application it is important that the Mermin inequality leads to a higher significance
than the Ardehali inequality, even if noise is introduced 
\cite{remarkimportant}.
To see this, we considered bit-flip noise, which can easily be 
simulated in experiment. Therefore, we
used a perfect GHZ state whose qubits are locally affected by the bit-flip operation $f$ with probability $p$, i.e.
$f(\varrho_{i}) =(1-p) \varrho_{i} + p X_i \varrho_{i} X_i$ for each qubit $i$.
In Fig.~\ref{fig:bitflipnoise}, we plotted the significance $\SSS$ versus the fidelity
$F$ of the noisy state w.r.t. a perfect GHZ state, i.e.
$F = \bra{{GHZ_4}} \varrho_{exp} \ket{{GHZ_4}}$, and versus the bit-flip probability $p$. For
$F \geq 0.70$ the Mermin inequality is more significant (for the 6-qubit versions of these inequalities \cite{Mermin, Ardehali}, this changes to $F \geq 0.40$). As can be seen from Eq.~(\ref{errorformula}), 
the fact that one witness is more significant than the other one is independent of the total particle number. Moreover, a calculation for white noise yields very similar values ($F \geq 0.72$ for 4 qubits, $F \geq 0.41$ for 6 qubits). This suggests that the effect does not depend on details of the noise. 
Note that the threshold value for white noise vanishes exponentially fast for an increasing number of qubits.

\begin{figure}
\includegraphics[width=0.8\columnwidth]{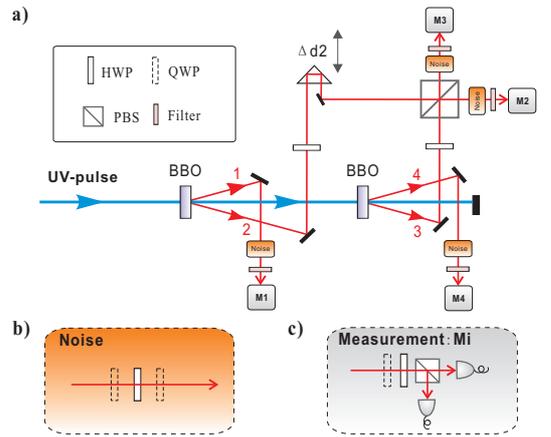}
\caption{\label{Bell} Scheme of the experimental setup.
\textbf{a}. The setup to generate the required four-photon 
GHZ state. Femtosecond laser pulses ($\approx$ 200 fs, 76 MHz, 
788 nm) are converted to ultraviolet pulses through a frequency 
doubler LiB$_{3}$O$_{5}$ (LBO) crystal (not shown). The pulses 
go through two main $\beta$-barium borate (BBO) crystals (2 mm), 
generating two pairs of photons. The observed two-fold coincidence 
count rates are about  $1.6 \times 10^{4}/$s with a visibility of 
96\% (94\%) in the $H/V$ ($+/-$) basis. \textbf{b}. Setup for engineering 
the bit-flip noise. \textbf{c}. The measurement setup.}
\end{figure}

%%%%%%%%%%%%%%%%%%%%%%%%%%%%%%%%%%%%%%%%%%%%%%%%%%%%%%%%%%%
{\it Experimental setup ---} 
%%%%%%%%%%%%%%%%%%%%%%%%%%%%%%%%%%%%%%%%%%%%%%%%%%%%%%%%%%
%Let us proceed with the experimental
%implementation. 
Spontaneous down conversion has been used to 
produce the desired four-photon state [see Fig.~\ref{Bell}(a)]. 
With the help of polarizing beam splitters (PBSs), half-wave 
plates (HWPs) and conventional photon detectors, we prepare a 
four-qubit GHZ state, 
%[see Eq.~(\ref{ghzdef})],
where
$\left| 0 \right\rangle = \left| H \right\rangle$ 
($\left| 1 \right\rangle = \left| V \right\rangle$) 
denotes horizontal (vertical) polarization.
%In the experiment, as a proof-of-principle,
We have chosen the bit-flip noise channel to demonstrate the theory
introduced in this paper. As shown in Fig.~\ref{Bell}(b), the noisy
quantum channels are engineered by one HWP sandwiched with two
quarter-wave plates (QWPs) \cite{chennoise}. The HWP is switched randomly between 
$+\theta$ and $-\theta$ and the QWPs are set at $0^\circ$ with respect 
to the vertical direction. In this way, the noise channel can be engineered with a bit-flip probability $p = \sin ^2 (2\theta)$. 
The Pauli matrix measurements required in the Bell test can be
implemented by a combination of HWP, QWP and PBS [see Fig.~\ref{Bell}(c)]. 
The fidelity of the prepared GHZ state is obtained via
$F = \frac{1}{2}(\left\langle {\left|
{0000} \right\rangle \left\langle {0000} \right| + \left| {1111}
\right\rangle \left\langle {1111} \right|} \right\rangle)  +
\frac{1}{{16}}\left\langle {\BB_{M}} \right\rangle$. Without added noise, its value is $F=0.84\pm 0.01$.

%%%%%%%%%%%%%%%%%%%%%%%%%%%%%%%%%%%%%%%%%%%%%%%%%%%%%%%%%%%%%%%%%%%%%%
{\it Experimental results ---} 
%%%%%%%%%%%%%%%%%%%%%%%%%%%%%%%%%%%%%%%%%%%%%%%%%%%%%%%%%%%%%%%%%5
For different noise levels, the experimental results of the violation, 
the statistical error and the significance are shown in Table I. The 
first observation is that, when there is no engineered noise, the 
violation of the Mermin inequality is smaller than the violation of 
the Ardehali inequality.
Its significance, however, is larger than that of the Ardehali inequality; this 
proves that testing the Mermin inequality is a better choice to characterize 
the entanglement in this case. Secondly, when the noise 
level increases, the significance in the Mermin inequality decreases more 
quickly. When $\theta=\pm 6^\circ, \pm 8^\circ$,
the significance for the Ardehali inequality is already larger than
that for the Mermin inequality. Due to the experimental imperfections, 
the initial state to which the 
%engineered 
noise is added is not the perfect GHZ state.
%\sout{Nevertheless, the theoretical predictions in Fig.~\ref{fig:bitflipnoise} are confirmed.}
However, assuming an initial state like
$
\varrho(p=0) = \alpha \ket{0000}\bra{0000} + \beta \ket{1111}\bra{1111}
+ \gamma (\ket{0000}\bra{1111}+\ket{1111}\bra{0000}) + \frac{\lambda}{16} \eins,
$
where $\alpha = 0.362, \beta = 0.522, \gamma = 0.398, \lambda = 0.12$ reproduces that for $p\leq 0.019 $ the Mermin inequality 
is more significant.

\begin{table}[t]
\begin{tabular}{||c|c||c|c|c||c|c|c||}
  \hline%\hline
  $\theta$       & p & $\VV(\mathcal{B}_M)$ & $\EE(\mathcal{B}_M)$ & $\SSS(\mathcal{B}_M)$ & $\VV(\mathcal{B}_A)$ & $\EE(\mathcal{B}_A)$ & $\SSS(\mathcal{B}_A)$ \\
  \hline
  $\pm 0^\circ$      & 0	 & 2.37 & 0.05 & 44.3 & 3.65 & 0.10 & 35.0 \\ \hline 
  $\pm 2^\circ$      & 0.005 & 2.00 & 0.06 & 33.4 & 3.14 & 0.11 & 29.2 \\ \hline
  $\pm 4^\circ$      & 0.019 & 1.57 & 0.07 & 23.7 & 2.48 & 0.11 & 21.8 \\ \hline
  $\pm 6^\circ$      & 0.043 & 1.13 & 0.07 & 16.2 & 2.05 & 0.11 & 17.8 \\ \hline
  $\pm 8^\circ$      & 0.076 & 0.67 & 0.08 & 8.8 & 1.63 & 0.12 & 13.7 \\
  \hline%\hline
\end{tabular}
\caption{Experimental values of the violation, the statistical error
and the significance for different values of $\theta$ (and the corresponding p). $\VV(\mathcal{B}_M)$,
$\EE(\mathcal{B}_M)$, $\SSS(\mathcal{B}_M)$ represent the values of $\VV$, $\EE$ and $\SSS$ 
in testing the Mermin inequality; $\VV(\mathcal{B}_A)$, $\EE(\mathcal{B}_A)$, $\SSS(\mathcal{B}_A)$ 
represent the corresponding values for the Ardehali inequality.
Each setting $X_1 X_2 X_3 X_4$ etc.~in the Mermin inequality 
is measured for 800 s, while each setting $X_1 X_2 X_3 A_4$ etc.~in 
the Ardehali inequality is measured for 400 s. The average total count 
number for each inequality is about 7500.}
\label{witnessmeasurent}
\end{table}

%%%%%%%%%%%%%%%%%%%%%%%%%%%%%%%%%%%%%%%%%%%%%%%%%%%%%%%%%
{\it Discussion ---}
%%%%%%%%%%%%%%%%%%%%%%%%%%%%%%%%%%%%%%%%%%%%%%%%%%%%%%%%%
%
We have proved that 
%in order to justify a claim of entanglement 
%in an experiment with a high statistical significance, 
it can be favorable to use an entanglement witness or 
a Bell inequality that results in a lower violation. 
We confirmed this experimentally using four-photon GHZ states. 
Our results show that the usual way of optimizing witnesses 
will not necessarily lead to more powerful 
tools for the analysis of many-particle experiments. It is important to 
note that when the number of photons in multi-photon experiments 
is increased the count rates decrease; consequently, the statistical 
error becomes more and more relevant.

Our results provide a direction to find powerful entanglement tests 
for low count rates: the observed effect relied on the fact that in 
the Mermin inequality only stabilizer measurements were made. There 
are already powerful approaches available to construct witnesses from 
stabilizer observables \cite{stabwit}  and also other Mermin-like or 
Ardehali-like Bell inequalities have been explored \cite{mermarde}. 
Consequently, these approaches are promising candidates for developing 
sensitive analysis tools. Further, inequalities similar to witnesses 
have been proposed and used to characterize quantum gate fidelities 
\cite{hofmann}, which is another application of our theory. 
Finally, we believe that results on statistical confidence from other 
fields of physics (e.g. \cite{feldman}) can give new insights in advanced
experiments on quantum information processing.

%Finally, we believe that a careful analysis
%of the statistical error can also help in designing loophole-free
%Bell tests.

We thank A. Cabello, R. Gill, O. Gittsovich, D. James, J.-\AA.  
Larsson and C. Roos for discussions and acknowledge support by the FWF 
(START Prize), the EU (SCALA, OLAQUI, QICS, Marie Curie Excellence Grant), the 
NNSFC, the NFRP (2006CB921900), the CAS, the ICP at HFNL, and the A.~v.~Humboldt 
Foundation.

%%%%%%%%%%%%%%%%%%%%%%%%%%%%%%%%%%%%%%%%%%%%%%%%%%%%%%%%%%%%%%
{\it Appendix ---} 
%%%%%%%%%%%%%%%%%%%%%%%%%%%%%%%%%%%%%%%%%%%%%%%%%%%%%%%%%%%%%%
To prove the Observation, we use as 
an ansatz for the improved witness $\WW'=\WW+\gamma P$,
where $\gamma > 0$ and $P$ is a positive observable 
with unit trace. For small $\gamma$, we expand
%\begin{multline}
$
-\frac{\langle \WW'\rangle}{\Delta(\WW')}
=-\frac{\langle \WW\rangle}{\Delta(\WW)}
+\gamma\frac{\langle \WW \rangle}{2\Delta^{3}(\WW)}\Bigl[\langle \WW P + P \WW \rangle
%\\
{}-2\frac{\langle \WW^{2}\rangle}{\langle \WW\rangle}\langle P \rangle \Bigr]+O(\gamma^2)\ .
$
%\end{multline}
Maximizing this expression over all positive $P$ with
$\tr(P)=1$ is equivalent to minimizing $\tr(Q P)$, where
$Q=\vr \WW+\WW\vr-2\langle \WW^{2}\rangle/\langle \WW\rangle\vr$.
Hence the optimal $P$ is a one-dimensional projector
$P=\ket{\varphi}\bra{\varphi}$, where $\ket{\varphi}$ is an
eigenvector corresponding to the  minimal eigenvalue of $Q$.
We still have to show that this minimal eigenvalue is
negative. %\cite{remarkappendix}. 
To this end, we make
the ansatz
$\ket{\varphi}=\alpha\ket{\psi}+\beta\ket{\psi^{\perp}}$,
where $\braket{\psi}{\psi^{\perp}} = 0$. We then have to minimize
%\begin{equation}
$
\tr(QP) = 2\real(\alpha^{*}\beta\bra{\psi}\WW\ket{\psi^{\perp}})
- 2\abs{\alpha}^{2}\frac{\Delta_\psi^{2}(\WW)}{\bra{\psi}\WW\ket{\psi}}
.
$
%\end{equation}
We can always choose the phases of $\alpha$ and $\beta$ such that
$\real(\ldots)$ is negative. Therefore the optimal $\ket{\psi^{\perp}}$
is the vector orthogonal to $\ket{\psi}$ which maximizes
$\abs{\bra{\psi}W\ket{\psi^{\perp}}}$, i.e.,
$\ket{\psi^{\perp}_{\text{opt}}}=[\id-\ket{\psi}\bra{\psi}]\WW\ket{\psi}/\Delta_{\psi}(\WW)$.
Furthermore, we can always choose the moduli of $\alpha$ and $\beta$ such that the negative
term $2\real(\ldots)$ dominates the positive second term. This shows that the minimal
eigenvalue of $Q$ is negative.

For finite $\gamma$ we can iterate this procedure. We always find
the same $\ket{\psi^{\perp}_{\text{opt}}}$ (though $\alpha$ and $\beta$
will be different in each iteration step). Thus, we make the ansatz $ \gamma P=a\ket{\psi}\bra{\psi}+b\ket{\psi^{\perp}_{\text{opt}}}\bra{\psi^{\perp}_{\text{opt}}}+c\ket{\psi}\bra{\psi^{\perp}_{\text{opt}}}+\text{h.\,c.} $ for the final result of the iteration. If we choose
$ c=-\Delta_{\psi}(\WW)$, $ab\ge\abs{c}^{2}$, and $a, b>0$,
then $\gamma P$ is positive, $\ket{\psi}$ is an eigenstate of $\WW',$
and $\Delta_{\psi}(\WW')$ is zero, so $\SSS$ diverges.
$\qed$


\begin{thebibliography}{15}
\bibitem{remarkidentical} Indeed, with the given choice of 
observables the
%Bell
operators $\BB_{\rm M}$ and $\BB_{\rm A}$ are identical within quantum 
mechanics.
However, the use of $A$ and $B$ within the Ardehali inequality results in a 
more restrictive test for LHV models.
\bibitem{remarkimportant}
This is also important, as for nearly perfect GHZ states, some count numbers 
$n_{kl}$ will be close to zero. Then, the interpretation of the statistical 
error as a confidence interval may be questioned.
\end{thebibliography}
\end{document}